\documentclass[iop]{emulateapj}
\usepackage{amssymb,amsmath,graphicx,psfrag,mathtools}
\usepackage{url}
\begin{document} 
\title{AR Sco: A Precessing White Dwarf Synchronar?} 
\shorttitle{AR Sco: A Precessing White Dwarf Synchronar?} 
\shortauthors{Katz} 
\author{J. I. Katz\altaffilmark{}}
\affil{Department of Physics and McDonnell Center for the Space Sciences} 
\affil{Washington University, St. Louis, Mo. 63130 USA} 
\email{katz@wuphys.wustl.edu}
\begin{abstract}
The emission of the white dwarf-M dwarf binary AR Sco is driven by the rapid
synchronization of its white dwarf, rather than by accretion.
Synchronization requires a magnetic field $\sim 100$ gauss at the M
dwarf and $\sim 10^8$ gauss on the white dwarf, larger than the fields of
most intermediate polars but within the range of fields of known magnetic
white dwarfs.  The spindown power is dissipated in the atmosphere of the M
dwarf, within the near zone of the rotating white dwarf's field, by magnetic
reconnection, accelerating particles that produce the observed synchrotron
radiation.  The displacement of the optical maximum from conjunction may be
explained either by dissipation in a bow wave as the white dwarf's magnetic
field sweeps past the M dwarf or by a misaligned white dwarf rotation axis
and oblique magnetic moment.  In the latter case the rotation axis precesses
with a period of decades, predicting a drift in the orbital phase of
maximum.  Binaries whose emission is powered by synchronization may be
termed synchronars, in analogy to magnetars.
\end{abstract}
\maketitle 
\section{Introduction}
\cite{M16,B16} recently discovered that AR Sco is a M dwarf/white dwarf binary
with an orbital period of 3.56 h and a white dwarf spin period of 1.95 m.
The white dwarf's spin is slowing with a characteristic time $P/{\dot P} =
0.9 \times 10^7$ y.  Most of the radiated power is emitted from the M dwarf
near the L$_1$ inner Lagrange point, but the phase of maximum is displaced
from conjunction by about 0.15 of the orbit.

AR Sco emits X-rays but they are only a small fraction of the system's
luminosity, implying that it is not powered by accretion.  Spindown is the
remaining plausible source of energy, with a power $- I \omega {\dot
\omega}$, where $I$ is the white dwarf's moment of inertia and $\omega$ its
rotation rate\footnote{Properly, $\omega$ is the WD rotation rate in a
rotating frame in which the total angular momentum of the binary is zero.
To an excellent approximation, this is the frame rotating at the orbital
rate, in which the WD spin rate is about 1\% less than in an inertial
frame.  The observed optical modulation is at this lower frequency.}.
The spindown power is converted to radiation with efficiency several times
the 2--3\% of $4\pi$ sterad subtended at the white dwarf by its companion.
This excludes reprocessing of roughly isotropic radiation from the white
dwarf as the source of the observed radiation.

I follow \cite{M16} in assuming the white dwarf's rotational energy is
dissipated by interaction of its magnetic field with the M dwarf's
atmosphere.  The spectral energy distribution and polarization \citep{B16}
indicate a substantial contribution of synchrotron radiation, implying that
the dissipation is by nonthermal processes, such as magnetic reconnection,
that accelerate energetic particles.

The rapid spindown suggests that AR Sco may be a missing link between
synchronously rotating polars (AM Her stars) whose synchronism is maintained
by magnetostatic interaction \citep{JKR79} to extreme accuracy \citep{K89}
and the asynchronous intermediate polars (IP).  Because AR Sco is still far
from synchronism the rate of dissipation of its white dwarf's rotational
energy is much greater than that in almost-synchronous systems like V1500
Cyg \citep{K91a,K91b}.

\begin{figure}
\centering
\includegraphics[width=3.3in]{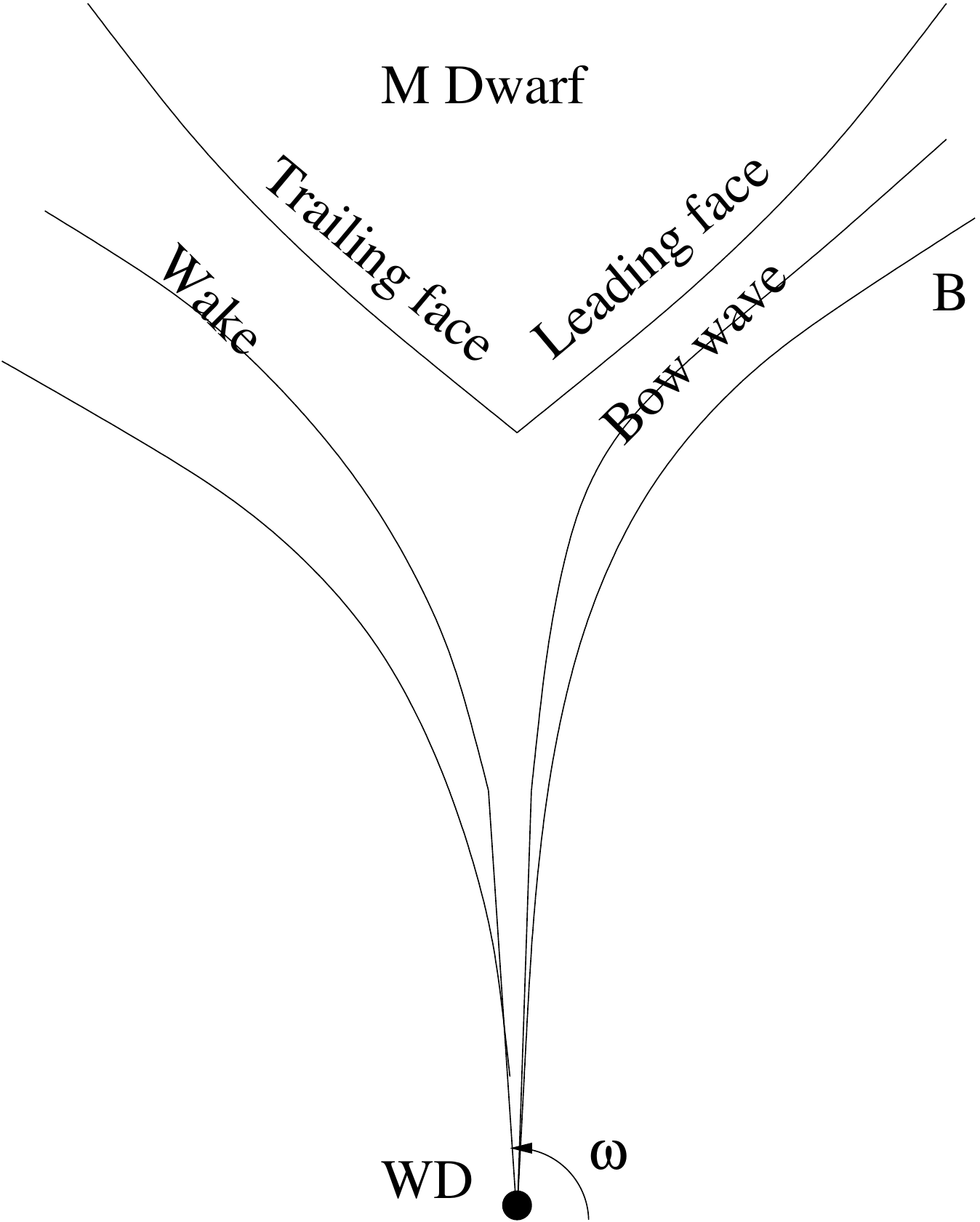}
\caption{\label{ARScoF1} The white dwarf's field is moving past the M dwarf
at a speed $a\omega$, where $a$ is the separation of the stars.  This is
fast enough (15--20\% of $c$) to induce an asymmetry in the magnetic
interaction between the ``upstream'' and ``downstream'' faces of the M
dwarf, even though there is not expected to be a particle wind from the
white dwarf.  The white dwarf's magnetic field is an electromagnetic wind,
even inside the radiation zone (though not deep in the near zone, close to
the white dwarf, where a magnetostatic treatment should be valid).  More
energy is released in a bow wave on the upstream side.  The Figure shows a
magnetic moment misaligned with the white dwarf's spin axis, but this is
not necessary; an upstream/downstream asymmetry and a torque would occur
even if the orbital, spin and magnetic axes were parallel, provided $a
\omega {\hskip 0.4mm \not \hskip -0.4mm \ll} c$.}
\end{figure}

\begin{figure}
\centering
\includegraphics[width=3.3in]{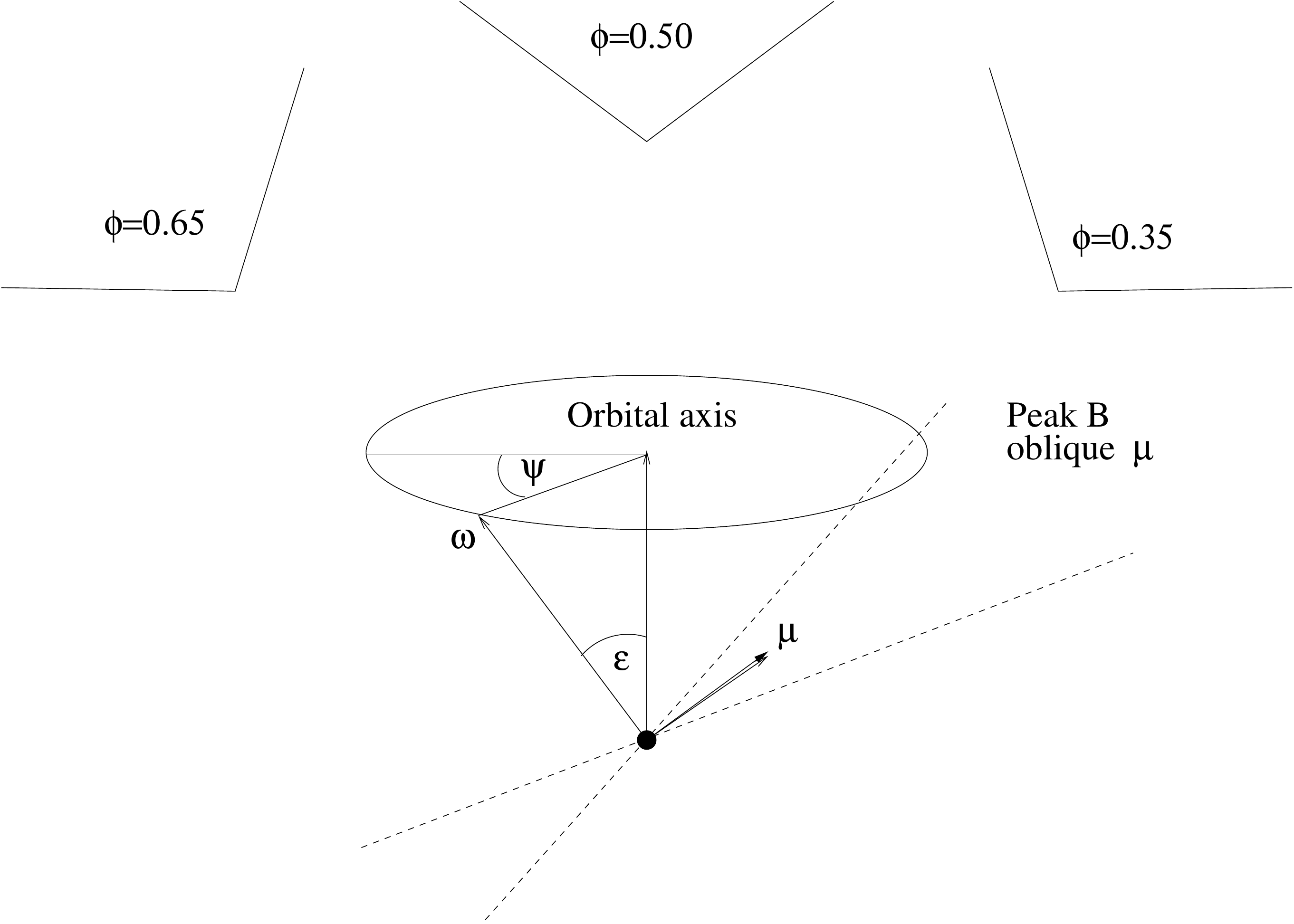}
\caption{\label{ARScoF2} The white dwarf's spin $\vec \omega$ makes an angle
$\epsilon$ with the orbital axis, and the spin axis precesses.  If the
magnetic moment is oblique (shown here as perpendicular) to the spin axis,
as indicated by the large amplitude modulation at the sideband of the spin
frequency, the magnetic field is maximum in a broad fan, indicated by dashed
lines, around the rotational equator, and dissipation is maximum at orbital
phases when the M dwarf (shown at three orbital phases) is in that fan.  In
the Figure that occurs at $\phi \approx 0.35$ and $\phi \approx 0.85$ (at
$\phi \approx 0.85$ the heated surface is not visible to the observer).}
\end{figure}

The visible brightness of AR Sco does not peak at conjunction (orbital phase
$\phi = 0.5$), when we might expect to see the heated side of the M dwarf
most fully, but rather around phase $\phi \approx 0.35$.  The purpose
of this paper is to investigate the origin of this surprising observation.
I consider two hypotheses:
\begin{enumerate}
\item The power dissipated by interaction between the white dwarf's magnetic
field and the M dwarf is greater on the latter's leading face where a
bow wave may form (Fig.~\ref{ARScoF1}) than on its trailing face.  The
dissipation rate does not depend on orbital phase but we view the hottest
portion of the M star more fully before conjunction.
\item The white dwarf's spin axis is not aligned with the orbital axis and
its magnetic moment is not aligned with its spin axis.  The dissipation rate
in the M dwarf's atmosphere depends on the orbital phase
(Fig.~\ref{ARScoF2}).  Precession of the spin axis makes the orbital phase
of maximum drift.
\end{enumerate}
In either hypothesis the optical maximum is displaced from conjunction.  In
the first hypothesis it precedes conjunction if the white dwarf spin is
prograde (and follows it if retrograde).  In the second hypothesis any
phase is possible, but precession makes it recede (if the spin is prograde).

\cite{GZH16} and \cite{B16} have suggested that the white dwarf acts as a
pulsar.  But the M dwarf is well within the near zone of the white dwarf's
dipole radiation field, where the non-radiative Poynting vector exceeds the
radiative Poynting vector by ${\cal O} (c/a \omega)^4 \sim 10^3$--$10^4$.
\cite{M16} also pointed out the difficulty of coupling an estimated 9\% of
the spin-down power, if radiated, to a M dwarf that occupies only 2--3\% of
the white dwarf's sky.  This again argues against mechanisms that radiate
the energy in a broad beam, such as that of a rotating dipole, and indicates
non-radiative coupling.

I therefore attribute the dissipation
to the interaction of the white dwarf's quasi-static vacuum magnetic field
with the M dwarf's atmosphere.  This interaction and the dissipation rate
vary with orbital phase, shifting maximum light from conjunction, even if
the white dwarf's rotation and magnetic axes are aligned (that alignment is
excluded in AR Sco because of the large amplitude modulation at the sideband
of the spin period).  The irregularly fluctuating brightness near the
orbital phase of maximum may be attributed to the magnetic storms
characteristic of magnetic reconnection.
\section{Magnetic fields}
The hypothesis that the white dwarf is spinning down as a rotating magnetic
dipole in vacuum \citep{GZH16,B16} requires magnetic fields greater than
known white dwarf fields.  Equating the spin-down power $- I \omega {\dot
\omega}$ to the radiated power $2 \mu^2 \sin^2{\theta} \omega^4/(3 c^3)$,
where $\theta$ is the angle between the spin and rotational axes, $I =
I_{50} \times 10^{50} \text{g-cm}^2$ is the white dwarf's moment of inertia
and its spindown rate ${\dot \omega} = - 1.8 \times 10^{-16}$ s$^{-2}$,
gives the magnetic moment
\begin{equation}
\mu = \sqrt{- 3 I {\dot \omega} c^3 \over 2 \sin^2{\theta} \omega^3} = 6.9
\times 10^{34} \sqrt{I_{50}} \csc{\theta}\ \text{gauss-cm}^3.
\end{equation}
The white dwarf's polar magnetic field
\begin{equation}
B_p = 2 {\mu \over R^3}.
\end{equation}
Numerical evaluation (Table~\ref{Bpolar}) yields fields $> 10^9$ G.  These
are outside the range of observed white dwarf fields and therefore
implausible.

\begin{table}
\centering
\begin{tabular}{|c|cc|}
\hline
$M/M_\odot$ & $\mu$ (gauss-cm$^3$) & $B_p$ (gauss) \\
\hline
0.88 & $1.5 \times 10^{35} \csc{\theta} $ & $1.1 \times 10^9 \csc{\theta}$ \\
1.08 & $1.0 \times 10^{35} \csc{\theta} $ & $1.6 \times 10^9 \csc{\theta}$ \\
1.22 & $6.8 \times 10^{34} \csc{\theta} $ & $2.5 \times 10^9 \csc{\theta}$ \\
1.30 & $4.3 \times 10^{34} \csc{\theta} $ & $3.6 \times 10^9 \csc{\theta}$ \\
\hline
\end{tabular}
\caption{\label{Bpolar} Implied white dwarf magnetic moments $\mu$ and
fields $B_p$ if the spindown of AR Sco is attributed to vacuum magnetic dipole
radiation by a magnetized white dwarf of the indicated masses.  Significantly
lower masses are excluded by orbital analysis \protect\citep{M16}.  White dwarf
parameters are taken from Table~\ref{WDdata}.}
\end{table}

If the spin-down power is dissipated magnetically in the M dwarf's
atmosphere the implied field is readily, but roughly, estimated.  Because
the parameter $a \omega/c \approx$ 0.15--0.20 is not very small, the 
quasi-static approximation \citep{JKR79} is inexact and the field on the M
dwarf's leading (with respect to the rotating magnetic field) face may be
substantially greater than that on its trailing face.  The resulting torque
\begin{equation}
I {\dot \omega} \sim {B_l^2 - B_t^2 \over 8 \pi} \pi R_M^2 a,
\end{equation}
where $R_M \approx 2.5 \times 10^{10}$ cm is the radius of an M5 dwarf,
$a= 6.0 \times 10^{10}(1 + M_M/M_{WD})\csc{i}$ cm is the separation of the
stars \citep{M16} and $B_l$ and $B_t$ are the magnetic fields on the leading
and trailing faces of the M dwarf, respectively.  Then
\begin{equation}
B \sim \sqrt{B_l^2 - B_t^2} \approx 60 \sqrt{I_{50} \sin{i}}\ \text{gauss}.
\end{equation}

The magnetic field at the surface of the white dwarf
\begin{equation}
B_{WD} \approx B \left({a \over R_{WD}}\right)^3 \sim 10^8\ \text{gauss}.
\end{equation}
A number of these parameters, especially $I$, are uncertain (over the
allowed range $0.81 M_\odot \le M_{WD} \le 1.29 M_\odot$ \citep{M16} $I$
varies by an order of magnitude; Table~\ref{WDdata}), but these fields are
plausible.  They point to an unusually (for a polar; \cite{W95}), but
not extraordinarily, magnetized white dwarf in AR Sco, consistent with its
rapid spindown and high rate of dissipation of rotational
energy.

\section{Inclination}
The inclination angle $i$ is not directly measured, but can be constrained
from the mass function and the upper bound on white dwarf masses.  The mass
function \citep{M16}
\begin{equation}
M_f \equiv {M_{WD}^3 \sin^3{i} \over (M_{WD} + M_M)^2} = 0.395 M_\odot,
\end{equation}
where $M_M$ is the mass of the M dwarf.  Substituting $M_{WD} \le 1.40
M_\odot$ and expanding in a Taylor series yields
\begin{equation}
\label{inclination}
\sin^3{i} \ge 0.42 + 0.49\left({M_M \over M_\odot} - 0.3\right) + \ldots,
\end{equation}
where it is assumed $(M_M/M_\odot - 0.3)$ is small.  Following \cite{M16},
we take $M_M = 0.3 M_\odot$.
\section{The Orbital Phase of Maximum}
There are two distinct explanations of the fact that the optical maximum of
AR Sco does not occur at conjunction, in contrast to expectation for a
classic reflection effect.
\subsection{Magnetic Bow Wave}
If the white dwarf were rotating slowly enough then a magnetostatic
treatment of its interaction with the M dwarf \citep{JKR79} would be valid.
There would still be dissipation \citep{K89,K91a,K91b}, likely enhanced by
turbulent (rather than just Ohmic) resistivity, but 
there would be symmetry between the leading and trailing faces of the M
dwarf and maximum light would occur at conjunction.

In AR Sco the parameter $\omega a/c \approx 0.15/\sin{i}$ is in the range
0.15--0.20, large enough to distinguish the magnetic interaction
between leading and trailing faces of the M dwarf (Fig.~\ref{ARScoF2}) and
to describe them as bow wave and wake.  Unlike the Solar wind, there is no
material wind from the white dwarf, only a vacuum electromagnetic field that
cannot be completely described by the magnetostatic near zone regime,
although it is well inside (by a factor of 5--7) the speed of light cylinder
and the radiation zone.
\subsection{Precession}
If the white dwarf's spin axis is not parallel to the orbital axis the
magnetic field at the M dwarf, and therefore the rate of dissipation, will
vary with orbital phase.  This provides a possible explanation for the fact
that the optical maximum does not occur at orbital phase 0.5, as it would
for a classical reflection effect.  A simplified possible geometry is shown
in Fig.~\ref{ARScoF2}, in which the magnetic axis is assumed to be
perpendicular to the rotation axis.  The maximum magnetic field, along the
magnetic poles of the white dwarf, then sweeps out the rotational equatorial
plane of the white dwarf.

The M dwarf crosses this plane twice per orbit.  If the dissipation rate
varies with the magnetic field this would introduce a dependence on orbital
phase.  The observed intensity would then be the result of a convolution of
this orbital phase dependence of the dissipation rate with the orbital phase
dependence of the visibility of the heated face of the M dwarf.  An
analogous result follows for any dependence of dissipation on the field
(for example, the dissipation might vary $\propto {\vec B} \cdot {\vec E}
\propto {\vec B} \cdot {\dot{\vec B}}$ or $\propto {\vec J} \cdot {\vec E}
\propto {\dot B}^2$, where Maxwell's equations imply $E \propto {\dot B}$.).

If the white dwarf's spin axis makes an angle $\epsilon$ with the orbital
axis, the spin axis will precess, causing the optical maximum to drift in
orbital phase.  This might explain the scatter \cite{M16} of brightness of
AR Sco at orbital phases 0.25--0.4, if the spin axis precessed during the
period of photometric observations.  The precession rate is
\begin{equation}
{\dot \psi} = {3 \over 2} {G M_M \over a^3} {I_3 - I_1 \over I_3}
{\cos{\epsilon} \over \omega},
\end{equation}
where $I_3$ and $I_1$ are moments of inertia about, respectively, the white
dwarf's polar and equatorial axes.  Using the orbital geometry,
\begin{equation}
a = {K_2 P \over 2 \pi \sin{i}}{M_{WD} + M_M \over M_{WD}} = {6.02 \times
10^{10} \over \sin{i}} {M_{WD} + M_M \over M_{WD}}\ \text{cm},
\end{equation}
where $K_2$ is the M dwarf's line-of-sight velocity amplitude, $P$ the
orbital period and $f_p$ an oblateness coefficient obtained from the
calculations of \cite{J64}.  The dynamic oblateness is\footnote{In several
places \cite{J64} omits the essential dimensional factor $B_{Chand}$
in his $A \equiv \omega^2 D/8 \pi G B_{Chand}$.}
\begin{equation}
\label{oblateness}
{I_3 - I_1 \over I_3} = f_p {8 \pi G B_{Chand} \over \omega^2 D},
\end{equation}
where $B_{Chand} = 1.96 \times 10^6$ g/cm$^3$ (for $\mu_e = 2$) and $D
\equiv y_0^{-2}$ The relevant properties of white dwarfs are summarized in
Table~\ref{WDdata}.

\begin{table}
\centering
\begin{tabular}{|c|ccccc|}
\hline
$M_{WD}/M_\odot$ & $R$ (cm) & $I$ (g-cm$^2$) & $y_0$ & $D$ & $f_p$ \\
\hline
0.40 & $1.1 \times 10^9$ & $1.2 \times 10^{51}$ & 1.29 & 0.6 & 72. \\
0.61 & $8.6 \times 10^8$ & $8.9 \times 10^{50}$ & 1.58 & 0.4 & 43. \\
0.88 & $6.5 \times 10^8$ & $4.7 \times 10^{50}$ & 2.24 & 0.2 & 31. \\
1.08 & $5.0 \times 10^8$ & $2.2 \times 10^{50}$ & 3.16 & 0.1 & 31. \\
1.22 & $3.8 \times 10^8$ & $9.7 \times 10^{49}$ & 4.47 & 0.05 & 31. \\
1.30 & $2.9 \times 10^8$ & $4.0 \times 10^{49}$ & 6.32 & 0.025 & 33. \\
\hline
\end{tabular}
\caption{\label{WDdata} Properties of zero temperature slowly rotating white
dwarfs \citep{C39,J64}.  A molecular weight per electron $\mu_e = 2$ is
assumed.  $y_0$ is Chandrasekhar's density parameter $y$ at the center of
the star ($y_c$ in \cite{J64}) and $D \equiv y_0^{-2}$.  The oblateness
coefficient $f_p$ is defined in Eq.~\ref{oblateness}.  A white dwarf,
especially one with a source of external heating (accretion, magnetic
dissipation) or with a hydrogen envelope, may have a photospheric radius
significantly in excess of the tabulated zero temperature radius $R$ for
homogeneous $\mu_e = 2$ composition, but its moment of inertia will be very
close to the tabulated value.}
\end{table}

We consider two limiting cases:  If $\sin{i}$ has the minimum value
permitted by Eq.~\ref{inclination} then
\begin{equation}
{\dot \psi} \ge {3 \over 2} {M_M M_f \over (M_{WD} + M_M)}
{f_p D \omega \over 8 \pi (K_2 P/2\pi)^3 B_{Chand}}\cos{\epsilon}\ s^{-1}.
\end{equation}
In the other limit $\sin{i} = 1$ and
\begin{equation}
{\dot \psi} = {3 \over 2} M_M \left({M_{WD} \over M_{WD} + M_M}\right)^3
{f_p D \omega \over 8 \pi (K_2 P/2\pi)^3 B_{Chand}}\cos{\epsilon}\ s^{-1}.
\end{equation}
Numerical results for AR Sco are shown in Table~\ref{results}.  The implied
precession periods are in the approximate range 20--200 y, and are short
enough that if precession is occurring it should be apparent in a decade of
data (for the longest periods) and perhaps in one or a few years (for the
shortest periods).  The condition $\sin{i} \le 1$ sets a lower bound of
about 0.81 $M_\odot$ on $M_{WD}$ \citep{M16}, so that the properties of
lower mass white dwarfs shown in Table~\ref{WDdata} are not relevant to AR
Sco.

\begin{table}
\centering
\begin{tabular}{|c|cc|}
\hline
$M_{WD}/M_\odot$ & $P_{precess}$ (minimum $\sin{i}$) & $P_{precess}$
($\sin{i} = 1$) \\
\hline
0.88 & 21 $\sec{\epsilon}$ y & 17 $\sec{\epsilon}$ y \\
1.08 & 50 $\sec{\epsilon}$ y & 30 $\sec{\epsilon}$ y \\
1.22 & 110 $\sec{\epsilon}$ y & 55 $\sec{\epsilon}$ y \\
1.30 & 200 $\sec{\epsilon}$ y & 100 $\sec{\epsilon}$ y \\
\hline
\end{tabular}
\caption{\label{results} Precession periods of white dwarf spin in AR Sco
for various white dwarf masses.  In all cases the M dwarf is assumed to
have a mass $M_M = 0.3 M_\odot$; the precession periods are roughly
inversely proportional to $M_M$ over its plausible range $0.2 M_\odot
\lessapprox M_M \lessapprox 0.4 M_\odot$.  $M_{WD} < 0.81 M_\odot$ is
excluded \protect\citep{M16}.}
\end{table}
\section{Discussion}
The remarkable properties of AR Sco support the suggestion of \cite{M16}
that it differs from most intermediate polars in that it is powered by
synchronization of the white dwarf spin with the orbital motion (a
``synchronar'') rather than by accretion.  The displacement of the optical
maximum from conjunction implies that energy is deposited asymmetrically on
the surface of the M dwarf.  This may be explained either as a consequence
of the subrelativistic phase speed of the rotating white dwarf 
magnetosphere at the M dwarf or of a white dwarf spin axis that is oblique
to the orbital axis.  In the former hypothesis the orbital phase of maximum
is likely to be fixed; in the latter hypothesis the spin axis precesses with
a period of decades, producing a drift in the orbital phase of maximum. 

In at least one other IP, AE Aqr \citep{IB08,M09}, synchronization is also
the chief source of energy and radio emission indicates particle
acceleration.  The photometry of AE Aqr is complicated by its flaring, but
the amplitude of the periodic modulation is low and is explicable as an
ellipsoidal variation without evidence of a reflection effect or other
interaction with the white dwarf \citep{vP89,B91}.  The secondary star of AE
Aqr is of earlier spectral type (K4V) than that of AR Sco (M5V) and likely
intrinsically more luminous, reducing the effect of magnetic heating on the
light curve.  AR Sco and AE Aqr may be members of a class of synchronars,
magnetic binaries whose emission is powered by the synchronization of their
compact (white dwarf) asynchronously rotating components.  AR Sco is an
extreme example in which the contribution of intrinsic stellar and
accretional luminosity is particularly small. 

Accretion contributes only a small fraction, perhaps none, of the luminosity
of AR Sco.  This is likely attributable to its strongly magnetized white
dwarf.  AR Sco also differs from non-accreting binary (neutron star) pulsars
with nondegenerate companions because the binary pulsars' companions are in
the radiation zones of the fast-spinning pulsars' magnetic fields, so
any interaction between the companion and the field does not couple back to
the pulsars' rotation.  In such a system pulsar spin energy may be deposited
in the companion, but the pulsar radiates in a broad dipole pattern.
This limits the coupling efficiency to no more than twice the companion's
subtended solid angle at the pulsar, inconsistent with observations of AR
Sco, and the pulsar despins as would an isolated object.  In contrast,
in AR Sco the coupling is near-zone, though not quite quasi-static, rather
than radiative.  Its efficiency is not limited by the subtended solid angle
and, as in a well-designed transformer, can approach 100\%.
\section*{Acknowledgements}
I thank T. R. Marsh for valuable discussions.

\end{document}